\PassOptionsToPackage{dvipsnames,svgnames,table}{xcolor}
\documentclass[preprint]{vgtc}               

\graphicspath{{figures/}{pictures/}{images/}{./}} 

\usepackage{times}                     

\usepackage{booktabs}                  
\usepackage{url}
\usepackage{multirow}
\usepackage{todonotes}
\usepackage{tabularx}
\definecolor{customgold}{HTML}{FFB612}
\definecolor{lightblue}{HTML}{A4B2F1}
\definecolor{yellow}{HTML}{F7D800}
\definecolor{grey}{HTML}{A6A6A6}

\DeclareRobustCommand*\circled[1]{\tikz[baseline=(char.base)]{
            \node[shape=circle,draw,inner sep=1.1pt] (char) {#1};}}

\usepackage{mathptmx}                  

\onlineid{1682}

\vgtccategory{Research}

\vgtcinsertpkg

\preprinttext{This is the author's version of the article. To appear in an IEEE ISMAR conference.}

\title{PersoNo: \underline{Perso}nalised \underline{No}tification Urgency Classifier in Mixed Reality}

\author{Jingyao Zheng\thanks{e-mail: jingyao.zheng@connect.polyu.hk}\\ %
        \scriptsize The Hong Kong Polytechnic University %
\and Haodi Weng\thanks{e-mail: wengvictor5@gmail.com}\\ %
     \scriptsize The Hong Kong Polytechnic University %
\and Xian Wang\thanks{e-mail: xiann.wang@connect.polyu.hk}\\ %
     \scriptsize  The Hong Kong Polytechnic University
\and Chengbin Cui\thanks{e-mail: chengbin.cui@connect.polyu.hk}\\ %
     \scriptsize  The Hong Kong Polytechnic University
\and Sven Mayer\thanks{e-mail: info@sven-mayer.com}\\ %
     \scriptsize TU Dortmund University 
\and Chi-lok Tai\thanks{e-mail: andy.tai@cpce-polyu.edu.hk}\\ %
     \scriptsize The Hong Kong Polytechnic University
\and Lik-Hang Lee\thanks{Corresponding Author. e-mail: lik-hang.lee@polyu.edu.hk.}\\ %
     \scriptsize The Hong Kong Polytechnic University
}

\abstract{
Mixed Reality (MR) is increasingly integrated into daily life, providing enhanced capabilities across various domains. However, users face growing notification streams that disrupt their immersive experience. We present \textit{PersoNo}, a personalised notification urgency classifier for MR that intelligently classifies notifications based on individual user preferences. Through a user study (N=18), we created the first MR notification dataset containing both self-labelled and interaction-based data across activities with varying cognitive demands. Our thematic analysis revealed that, unlike in mobiles, the activity context is equally important as the content and the sender in determining notification urgency in MR. Leveraging these insights, we developed \textit{PersoNo} using large language models that analyse users' replying behaviour patterns. Our multi-agent approach achieved 81.5\% accuracy and significantly reduced false negative rates (0.381) compared to baseline models. \textit{PersoNo} has the potential not only to reduce unnecessary interruptions but also to offer users understanding and control of the system, adhering to Human-Centered Artificial Intelligence design principles.
}

\begin{document}


\firstsection{Introduction}

\maketitle

Mixed Reality (MR) environments are increasingly integrated into daily life, blending digital information with physical surroundings. In this paper, we treated MR as synonymous with Augmented Reality: virtual objects integrated into the real world~\cite{speicher2019mixed}. It enhances human capabilities across manufacturing~\cite{gonzalez2017immersive} and education~\cite{maas2020virtual}. However, to avoid losing touch with reality, users face growing notification streams in MR, which present unique challenges as they distract users from their immersive experience with no task-related information. Virtual Reality (VR) studies have discussed similar concerns of breaking the immersive experience and emphasised the importance of notifications~\cite{ghosh2018notifivr, hsieh2020bridging, rzayev2019notification}. Yet isolating users from notifications induces anxiety and disconnection~\cite{pielot2017productive}. This contradiction underscores the need for intelligent MR notification classifiers to filter notifications and cause less disruption appropriately.

\begin{figure}[t]
  \centering
  \includegraphics[width=\linewidth]{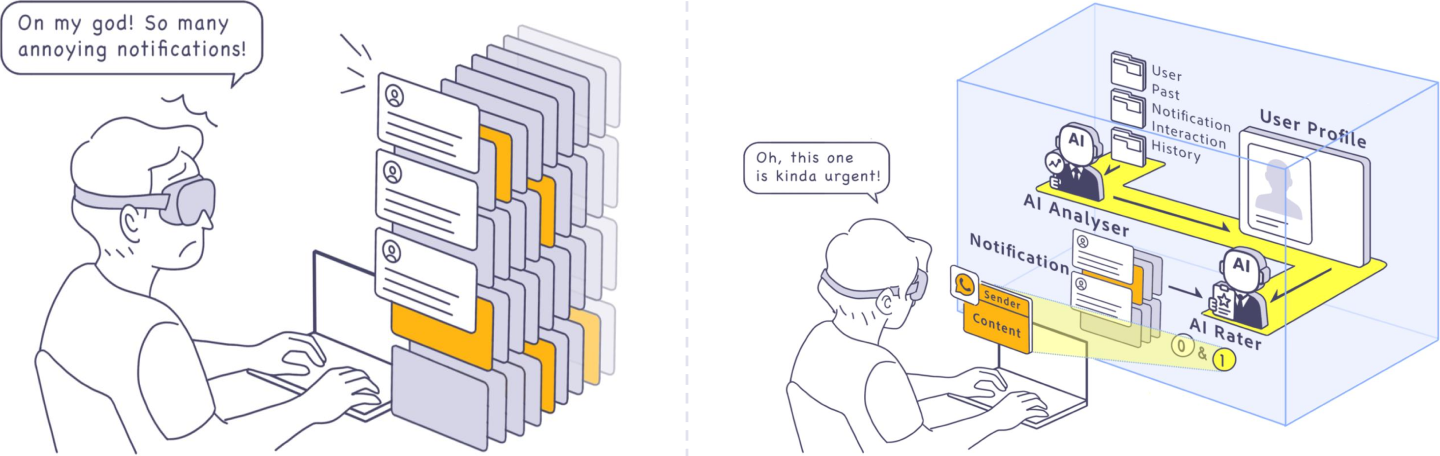}
  \caption{\textbf{Left}: Users are overwhelmed by MR notifications during work, with only a few {\color{customgold} urgent notifications}. \textbf{Right}:  With \textbf{PersoNo}, only urgent notifications are pushed to the user. The {\color{lightblue} box} illustrates the system {\color{yellow} workflow}—the AI analyser processes the user's past notification interaction history to generate user profiles, which AI raters then use to evaluate upcoming notifications.} 
  \label{fig:teaser_figure}
\end{figure}

Human-computer interaction (HCI) researchers have studied how to balance user attention and interruptions in mobile settings for a long time. Prior work has revealed that the content of a message is a primary factor determining its urgency and a user’s likelihood of responding~\cite{li2023alert, mehrotra2015designingContentDriven}. In addition, users' recent interactions with senders allow them to make accurate notification speculations~\cite{chang2019think}, indicating the sender is pivotal in determining users' receptivity to the notifications. Prior approaches~\cite{mehrotra2015designingContentDriven, oh2015intelligent} have leveraged such insights to develop intelligent context-aware smartphone notification systems. However, mobile users can easily distance themselves from interruptions by putting their phones away. In contrast, to truly escape notifications in MR, users must completely remove their head-mounted devices (HMDs), which not only breaks the immersive experience but also eliminates access to all the augmented capabilities that make MR valuable for their work.

Unlike mobile notification researchers, MR/VR researchers primarily focused on notification design and multimodal approaches~\cite{ghosh2018notifivr, marques2022which, plabst2025order, rzayev2020effects}. Limited research exists on intelligent notification management~\cite{chen2022predicting, lindlbauer2019context}, yet we argue it is crucial for alleviating user distraction. To date, no established dataset or framework captures how users handle incoming notifications in MR, leaving developers without datasets for developing MR notification classifiers. This knowledge gap, coupled with the practical importance of minimising user distraction in MR, motivates our research.

In this work, we aim to manage notifications intelligently to reduce user distraction while maintaining important notification updates. We address two research questions that drive our study and facilitate the development of the Personalised Notification Urgency Classifier in MR (PersoNo): \textbf{RQ1: How do users behave and respond to notifications in MR?} We seek to understand the human side of the problem: when an MR user receives a notification, what factors influence whether they attend to it or ignore it? Would the variables be the same as those in mobile notification interaction? These responses inform the key variables to be considered in developing MR notification classifiers, providing critical insights into contextual and user-specific factors. Regarding the second research question, we draw upon previous notification classifier research~\cite{chen2022predicting, dredze2008intelligent, li2023alert}, which demonstrated that personalised models trained exclusively on individual user data outperform general models trained on aggregated multi-user datasets. Based on these findings, we formulate \textbf{RQ2: How can we automatically classify the urgency of MR notifications in a personalised manner?} which encompasses three main dimensions for developing a classifier: Data, Context, and Algorithm.

To address these questions, we conducted a detailed study and developed a solution with three key contributions, corresponding to three PersoNo essential elements (Data, Context and Algorithm): \textbf{(1)} We created a \textbf{new MR notification dataset} through a user study (N=18) where participants wore MR headsets while experiencing everyday tasks and received messages. We collected objective and subjective data through self-labelling (users' rating notification urgency) and by tracking actual response behaviours. With this first-of-its-kind dataset ($N=18 \times 198$), we demonstrated that self-labelling offers a convenient alternative to activity-based data collection for future PersoNo deployment, yielding comparable classifier performance. \textbf{(2)} We analysed \textbf{users' replying behaviour patterns regarding MR notifications}. Our findings revealed certain patterns consistent with mobile research. For example, message content emerged as a critical factor when determining whether to attend to notifications~\cite{li2023alert}. However, we also discovered MR-specific insights. Notably, activities were reported with similar frequency as content when users described their behavioural patterns. \textbf{(3)} Our proposed \textbf{PersoNo} algorithm leverages Large Language Models (LLMs) and its classifier could accurately predict notification urgency by analysing users' replying behaviour patterns in small notification datasets.

\section{RELATED WORK}
\paragraph{Digital Notifications}
Researchers have examined notifications on smartphones and other personal devices. Mobile users receive dozens of notifications per day (around 63.5 on average), primarily from messaging and email, and typically attend to them within minutes due to social pressures~\cite{pielot2014situ}. While frequent alerts can induce stress or a sense of interruption, users also report feeling more connected when messaging notifications keep them aware of social updates~\cite{chang2023not}. Notably, complete avoidance of notifications is not a viable solution; experiments disabling push alerts found that users experienced anxiety and isolation without these ambient cues~\cite{pielot2017productive}. This underscores the need to manage rather than eliminate digital interruptions. Prior work identified key factors that determine which notifications users deem urgent or worthy of immediate response. The content of the message is consistently found to be a primary influence on perceived urgency and responsiveness~\cite{li2023alert, mehrotra2015designingContentDriven}. For example, critical or work-related content demands quicker attention than trivial updates. The sender is another pivotal factor: Chang et al.~\cite{chang2019think} observed that users often speculate about who a notification is from, and recent interactions with a sender strongly influence whether they will check the alert immediately. Other contexts also influence users' receptivity. These contexts include location~\cite{mehrotra2015designingContentDriven, pejovic2014interruptme}, time of day~\cite{poppinga2014sensor, sarker2014assessing} and activity context~\cite{aminikhanghahi2017thyme, mehrotra2016myPhoneAndMe}. These insights informed the design of intelligent notification management systems that attempt to filter or rank alerts by importance.

While existing notification research provides valuable insights for users' receptivity to mobile notifications, these findings may not translate to MR directly. Unlike mobile notifications that users can physically distance themselves from, MR notifications are inherently more invasive due to their immersive presentation within the user's field of view. Users must either endure disruptions or remove headsets entirely, sacrificing all augmented capabilities. It might potentially increase users' notification fatigue. This fundamental difference necessitates specialised approaches for MR notification management. Our work investigates how established factors influencing notification receptivity manifest differently in MR, and develops personalised intelligent systems based on the most significant contexts.

\paragraph{Personalised Notification Classifier}
Research on intelligent notification systems has explored multiple strategies: opportune time predicting~\cite{chen2022predicting, okoshi2017attention, pradhan2017understanding, sarker2017designing} and notification management~\cite{li2023alert, mehrotra_prefminer_2016}. Building upon this foundational research, subsequent studies~\cite{chen2022predicting, dredze2008intelligent, li2023alert} have compared models trained on personal and generic data, revealing a consistent pattern: personal data enables higher accuracy in notification classification. This raises a critical research challenge: how to construct an intelligent management system with limited training data.

Prior works developed intelligent notification management systems from both subjective~\cite{chen2022predicting, mehrotra2016myPhoneAndMe} and objective experience~\cite{mehrotra_prefminer_2016, pielot2017beyond}. Inspired by this, we compared two data collection methods: self-labelling and interaction, both previously used in message classification research~\cite{dredze2008intelligent, chen2022predicting}. They were included in our study, as each offers distinct advantages. Previous interaction-based data collection typically required several weeks to gather sufficient classifier training data. For example, Mehrotra et al.~\cite{mehrotra_prefminer_2016} needed a 15-day experiment to collect mobile notifications, while Pielot et al.~\cite{pielot2017beyond} spent an average of four weeks gathering data. In contrast, self-labelling allows notifications to be categorised within a much shorter timeframe. This efficiency could significantly enhance user acceptance of classifier applications, as users typically prefer applications that are ready for use shortly after deployment~\cite{maister1984psychology}. However, prior work~\cite{dang2020self} indicates that self-reporting and actual behaviours are only weakly correlated. It suggests that self-labelled data might not be reliable for training personalised classifiers. Our study, therefore, aims to compare the accuracy of classifiers trained on both types of data to determine whether self-labelled data can effectively substitute for interaction-based data in this context.

\paragraph{Notifications in Mixed Reality}
As computing extends into immersive environments like VR, notification management faces new challenges. In VR, users can become so engrossed that they miss critical external messages, leading to frustration when important information is delayed~\cite{hsieh2020bridging}. To address this challenge, numerous VR researchers have investigated optimal placement strategies to ensure user visibility and attention~\cite{hsieh2020bridging, rzayev2019notification, imamov2020display}. Besides the visual cues, Ghosh et al.~\cite{ghosh2018notifivr} explored multiple modalities for VR notifications. Among visual, aural, and haptic notifications, haptic ones were the least effective, a finding that aligns with subsequent research by George et al.~\cite{george2020invisible}.

While VR notification research has made considerable progress, MR introduces further complexity as digital information overlays the real world rather than replacing it. Prior works have explored the effectiveness of notifications across different multimodalities~\cite{marques2022which, cho2025evaluating}. From a visual perspective, Rzayev et al.~\cite{rzayev2020effects} demonstrated that notification positioning significantly impacts user perception, with proper alignment crucial for minimising distraction while maintaining awareness. Notably, Li et al.\cite{li2024predicting} developed a computational framework that predicts virtual element noticeability by analysing visual saliency patterns to anticipate when users detect element changes.

While existing research in MR and VR has predominantly focused on notification design elements like placement, modality, and visual appearance, there remains a significant gap in an intelligent MR notification management system. In this context, early work by Orlosky et al.~\cite{orlosky2014managing} showed that using see-through HMDs to relay mobile notifications can increase message awareness with minimal performance impact compared to checking handheld phones. However, this advantage diminishes in high cognitive load situations where users exhibit varying receptivity to interruptions; only HMDs equipped with intelligent notification systems that adapt to users' cognitive states and contextual preferences can truly deliver benefits without compromising task performance, such as the adaptive MR user interfaces based on users' cognitive load~\cite{lindlbauer2019context}. Our research first investigates the most significant contexts affecting MR users' receptivity to notifications, leveraging participant-reported contexts to develop an MR notification management system that enhances user experience.  More broadly, our work focuses on mitigating distraction in MR. While previous research has explored distraction reduction in MR/VR generally~\cite{Distraction_ISMAR1, Distraction_ISMAR, Distraction_ISMAR2}, our work makes a distinct contribution by specifically focusing on distraction mitigation through intelligent notification management.

\begin{figure}[t]
    \centering
    \includegraphics[width=\linewidth]{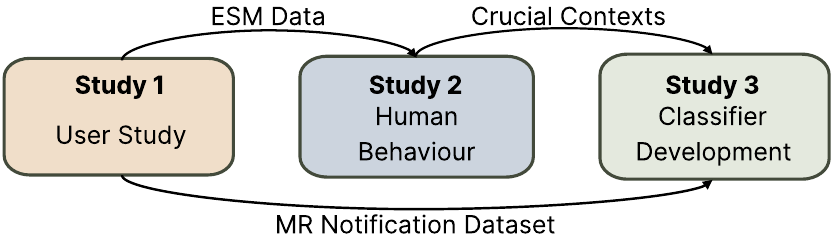}\caption{The research approach we took for our investigation.}
    \vspace{-5mm}
    \label{fig: workflow}
\end{figure}

\section{Research Approach}

Our research addresses interaction challenges in MR notification systems through a three-stage approach depicted in \autoref{fig: workflow}. Each study addressed one core PersoNo classifier component: Study 1 focused on Data, Study 2 on Context, and Study 3 on Algorithm. Study 1 collected an MR notification dataset through a user study, complemented by the Experience Sampling Method (ESM)~\cite{hektner2007experience} to capture participants' behavioural patterns through immediate self-reports. This facilitates subsequent analysis of optimal data collection methods (self-labelled or interaction-based) for future PersoNo deployment. The ESM data directly informed Study 2, where we analysed human behaviour patterns regarding notification interaction in various contexts. This identifies the key variables PersoNo should consider when classifying notification urgency levels. Building on these insights, Study 3 leveraged the MR notification datasets and crucial contexts to develop an intelligent MR notification management system, PersoNo. 

Our work identifies crucial contextual variables that influence notification receptivity in MR. It provides both empirical insights into user behaviour and practical solutions for reducing notification distraction while maintaining awareness of important information. This integrated approach bridges human-centred research with advanced Artificial Intelligence (AI) techniques to address a significant usability challenge in emerging MR interfaces.

\section{Study 1: MR Notification Data Collection}
\label{Label Collection}
To the best of our knowledge, the field lacks a comprehensive MR notification interaction dataset. To build an MR notification classifier, our research requires a user study to collect MR notification data. This collection serves two key purposes: analysing how participants respond to notifications in MR and developing an effective notification classification system.

To collect both the subjective and objective notification dataset, two phases (self-label phase and interaction phase) were conducted in a counter-balanced order. In the self-label phase, participants assessed the urgency levels of 90 randomly selected notifications from our dataset (details in Section~\ref{subsec:notification_dataset}), given the message content and the senders. For the interaction-based data collection phase, participants engaged in three MR activities, each consisting of two ten-minute sessions. Between sessions, there was at least a one-minute break. Our system recorded participants' behavioural responses and classified notifications as `non-urgent' when participants either ignored or dismissed them, and as `urgent' when participants actively chose to respond within 30 seconds.

Our approach classified notification urgency into two categories: \textit{urgent} and \textit{non-urgent}. This binary classification builds upon the work of Weber et al.~\cite{weber2019annotif}, who initially identified four notification clusters (\textit{C1}, \textit{C2}, \textit{C3}, and \textit{C4}) in daily interactions. Their research revealed that only \textit{C1} notifications demanded immediate user attention, while \textit{C2}, \textit{C3}, and \textit{C4} could be addressed at the user's convenience. Thus, we used a binary classification in our study based on whether immediate attention is required, which also aligns with the previous notification research design~\cite{mehrotra2015designingContentDriven}. We define the \textit{urgent} notifications as those that require replies within 30 seconds, and \textit{non-urgent} notifications which did not have this time constraint.

We initially hypothesised that considering only two key variables (content and sender) could achieve high prediction accuracy, as previous research~\cite{chang2019think, li2023alert, mehrotra2016myPhoneAndMe} suggested that these variables alone could yield reasonable results. Additional factors incorporated into the further analysis include the activities in which users were involved and their established messaging reply habits, as determined through the following thematic analysis.
Our contribution in this section lies in the construction of an MR notification dataset based on users' interaction behaviour during the MR activities ($N=18\times108$, comprising 18 participants with 108 notification data points per participant) and a self-labelled dataset ($N=18\times90$). 

\subsection{Mixed Reality Notification}

\subsubsection{Notification Dataset}
\label{subsec:notification_dataset}

Our study focused exclusively on instant messaging (IM) notifications, as mobile IM messages are anticipated to become a fundamental MR component~\cite{lee2023exploringcscw}. We selected WhatsApp as the application source due to its widespread use in our region. For the MR activity data, we carefully balanced the quantity of data with notification frequency. While aiming to maximise data collection, we avoided pushing notifications too frequently to prevent user annoyance, establishing a reasonable notification interval (See Section~\ref{subsub: DataCollection}). In total, we collected 108 notification instances during the MR activities and separated them into training and testing datasets (90 and 18, respectively; more details in Subection~\ref{sub:Classifier/Data Preparation}). Our approach follows established methodological practices in notification research~\cite{chen2022predicting, li2023alert} that separate activity-generated data into distinct training and testing datasets. To ensure equivalent training sets across both collection methods, we also gathered 90 self-labelled data, resulting in 198 notification data.

To protect participant privacy, we used Python scripts to randomly extract 198 data from the online Mobile Text Dataset (\textit{mobile\_train.txt})~\cite{vertanen2021mining}. This dataset is grounded in real-world mobile user behaviour. Originally, the dataset only contained the message content. To emulate the real-world experience, we assigned sender placeholders for each notification, such as \textit{friend 1} and \textit{friend 2}. Similar to the previous work~\cite{rzayev2019notification}, we collected the names of participants' friends and supervisors to replace the sender placeholders before the study and used these names as message senders during the experiment. To further enhance the realistic experience, we also incorporated group messages at proportions similar to those reported in Pielot et al.'s  work~\cite{pielot2018dismissed} (40 group messages and 158 messages).

\subsubsection{Notification Interaction}

\begin{figure}[b]
    \vspace{-4mm}
    \centering
    \includegraphics[width=\linewidth]{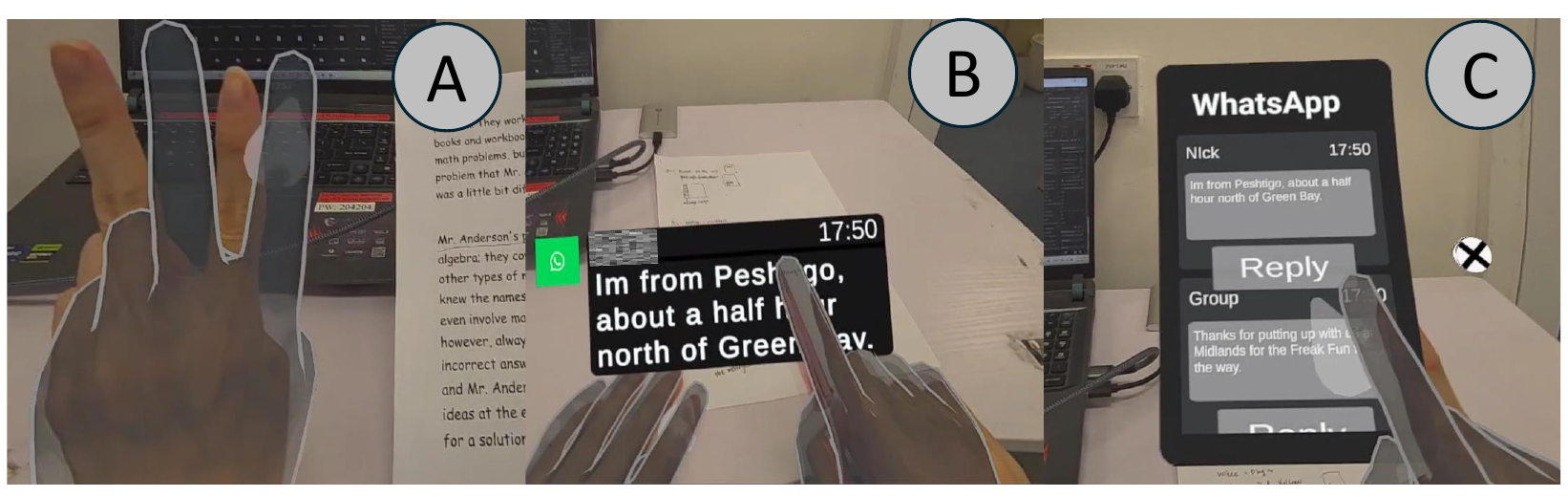}
    \caption{Examples of Notification Interaction. \circled{A} demonstrates the gesture for dismissing notifications in MR. \circled{B} illustrates the tapping interaction to access the notification panel displayed in \circled{C}, which features the ``Reply'' button that initiates the response workflow.}
    \label{fig: notification interaction}
\end{figure}

We designed notification interaction to mirror actual behaviours: users can ignore, actively dismiss, or respond to notifications (see \autoref{fig: notification interaction}). During the activities, participants may ignore or dismiss \textit{non-urgent} notifications. If they did not actively dismiss or respond within 20 seconds, notifications were automatically dismissed and stored in the notification panel (see \autoref{fig: notification interaction}C), aligning with previous research~\cite{rzayev2019notification}. Additionally, participants had the option to manually dismiss notifications using a specific gesture (see \autoref{fig: notification interaction}A). A response was required for notifications deemed urgent by the participants. Unlike the prior studies \cite{chen2022predicting, rzayev2019notification} that only allowed quick responses through simple controller presses, our study required participants to take additional steps to reply due to the common practice where users typically respond to notifications manually rather than using the smart reply~\cite{kannan2016smart}. 

Participants needed to open the app by either tapping the notification or gesturing to respond to notifications. We omitted the message-typing step and utilised the ``Reply'' button click to simulate the reply process (see \autoref{fig: notification interaction}C) to streamline the process and avoiding fatigue and dizziness caused by longer study duration. 

\subsubsection{Notification Display}

We adopted a notification user interface design (see \autoref{fig:notification}D-F) similar to previous work~\cite{chen2022predicting, rzayev2019notification}, which displays the sender, an image of the application source, and the content. All notifications were placed within the participants' field of view and designed to be easily noticeable.  Specifically, notifications were placed in the bottom centre of the user's field of view, as prior work~\cite{rzayev2018reading} showed this position improves comprehension and reduces distraction while sitting and Plabst et al.~\cite{plabst2022push} found subtitles provide higher comprehension and noticeability than heads-up displays. We positioned notifications 0.25 meters from users, closer than the Quest 3's focal distance of over 1 meter, to accommodate table-based MR tasks. Greater distances risked users reaching through the table when tapping, potentially causing injury. Overall, notifications were placed 0.25 meters away and angled 25° below the user's line of sight. 

\subsection{Procedure}

The entire study lasted approximately two hours. Upon arrival, participants were welcomed and provided with the information sheet detailing the study's purpose. Then, they signed a consent form and completed a demographic questionnaire. As mentioned earlier, before conducting the formal study, we asked their friends and bosses/supervisors for a few names. This information was filled in our notification dataset to replace the placeholders, such as \textit{Friend 1} and \textit{Supervisor}. Then, we utilised Python scripts to randomly separate the data into two groups: self-labelled notifications (90) and MR activity notifications (108). Following this, we counterbalanced the order of the two data collection methods. During the self-label phase, participants were asked to carefully read the notification content and the sender, and then rate the urgency levels of each notification according to their preferences and daily habits. The self-labelling session was conducted on a laptop. Regarding the MR activity part, we conducted an introductory session using slides to outline basic interactions with notifications and the primary tasks for each activity. Additionally, we developed an introductory VR scene that allowed participants to familiarise themselves with the notification interactions in a controlled environment. Once they were confident in the interaction, the MR activities were conducted in a Latin Square order to alleviate the carryover effects.

\subsection{Participants}
\label{sub:participant}
Our university's ethics board approved the study. We compensated participants through course credit or payment at the local minimum wage. We recruited 18 participants (5 males and 13 females) from our universities, aged from 18 to 27 years (M = 22.61, SD = 2.63). All participants had either normal or corrected-to-normal vision and were able to view notification details clearly. To measure participants' familiarity with MR, we used a 5-point Likert scale, where 1 indicated very inexperienced (``I have only used MR once or twice before, if at all'') and 5 indicated very experienced (``I use MR several times a month''). Results showed that participants were generally unfamiliar with MR (M = 2.5, SD = 1.12).

We also collected data regarding participants' mobile IM notification preferences. Participants were asked to select contexts and the most important one affecting their receptivity to mobile IM notifications, including Notification Content, Sender, Application Source, Cognitive Load, Time, Location, Mood, Activity, and others~\cite{chang2023not, mehrotra2016myPhoneAndMe}. 14 participants identified Notification Content as a significant context, while 7 selected Sender and 9 chose Activity. Regarding the most influential factor, Notification Content emerged as the most influential factor, reported by 10 out of 18 participants, followed by Sender (N=3) and Application Source (N=3). Individual contextual factors such as Mood (N=1) and Activity (N=1) showed notable variability among participants. These findings align with previous research indicating that most participants consider notification content the most influential contextual factor~\cite{li2023alert}.

\subsection{Experiment Design}
\paragraph{Mixed Reality Activities}
\begin{figure}[t]
    \centering
    \includegraphics[width=\linewidth]{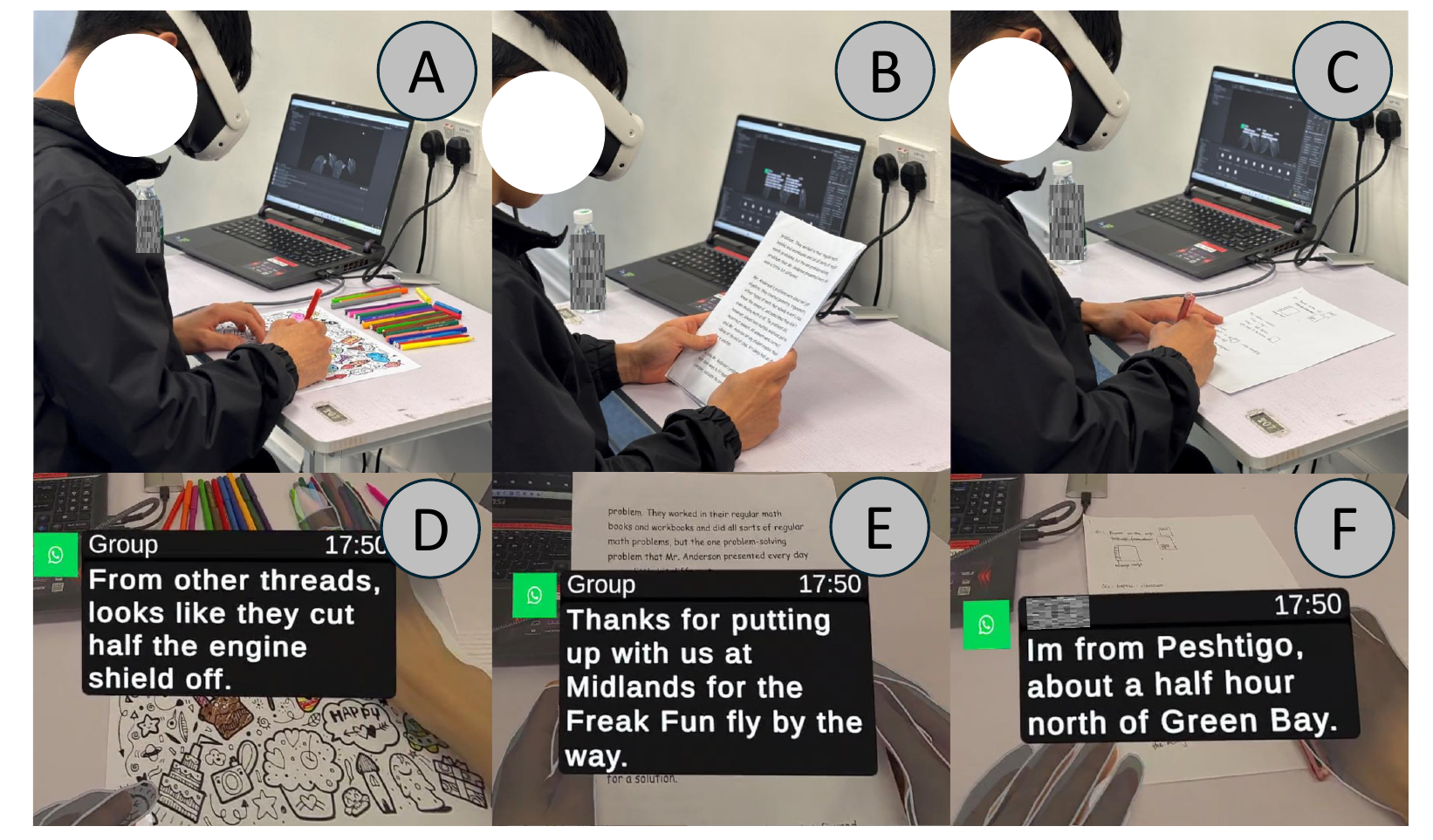}
    \caption{Examples of Mixed Reality Activities. \circled{A} shows the \textit{Doodling} Activity. \circled{B} illustrates the \textit{Reading and Comprehension} Activity. \circled{C} presents the \textit{Brainstorm} Activity. \circled{D}, and \circled{E}, \circled{F} show the examples of notifications received during the activity.}
    \vspace{-5mm}
    \label{fig:notification}
\end{figure}

Users engaged in three MR activities: \textit{Doodling}, \textit{Brainstorming}, and \textit{Reading and Comprehension} (see \autoref{fig:notification} (A-C)). These activities require varying levels of cognitive load, ranging from low to high~\cite{lindlbauer2019context}. It created a realistic testing environment where users encountered notifications across different mental states and activity types, similar to daily life experiences. 
During the \textit{Doodling} activity, participants were provided with five different plain graffiti drawings and coloured pens. They were free to choose one drawing and doodle without specific requirements. For the \textit{Brainstorming} activity, participants focused on designing future notifications in MR. Following Rietzschel et al.'s guideline~\cite{rietzschel2006productivity}, participants wrote their ideas on A4-sized sheets of paper and were encouraged to think creatively without concerns about feasibility. For example, they were prompted to consider multi-modal notifications beyond visual elements, including haptic feedback and taste. The \textit{Reading and Comprehension} materials and questions were sourced from \href{https://www.easycbm.com/}{easyCMB}~\cite{alonzo2006easycbm}, a resource widely used in previous research~\cite{chen2023easyCBMCHI, joshi2024easyCBMCHI}. Participants were instructed to read as quickly as possible while ensuring accuracy in their answers.

\paragraph{Data Collection}
\label{subsub: DataCollection}
The primary objective of our experiment was to collect notification data categorised by urgency levels. The study consisted of two parts: a self-label session and an interaction session. In the self-label session, participants labelled the urgency levels of each notification (90 notifications) in a CSV file. In the interaction session, participants engaged in three MR activities, with each activity divided into two 10-minute sessions. For each activity session, participants performed a primary task (the MR activity) while simultaneously handling a secondary task: reading notifications carefully and deciding whether to reply to messages. During each 10-minute activity session, 18 notifications were sent to participants at random intervals ranging from 20 to 32 seconds (cf. \cite{rzayev2019notification}). Our script automatically classified notifications based on response time: if a notification received a reply within 30 seconds, it was labelled as \textit{urgent} (1); otherwise, it was labelled as \textit{non-urgent} (0). Our system yielded two distinct datasets for analysis in Section~\ref{sec: Classifier}. Dataset 1 comprises information about the Sender, Content, and Urgency level, while Dataset 2 expands upon this by including Sender, Content, Urgency level, and Activity context. To further protect participant privacy, we parsed their friend and supervisor names to generic placeholders (e.g., \textit{Friend 1} and \textit{Supervisor}) throughout the dataset. 
To gain deeper insights into participants' decision-making processes, we applied the Experience Sampling Method (ESM)~\cite{hektner2007experience}. During the user study, at least one researcher periodically asked them to explain their notification response patterns. For example, a common prompt was: ``Could you please explain why you replied to your friend's last message?''. These qualitative insights were summarised and documented. It was incorporated into our analysis (see \autoref{sec:human_behaviour}).

Additionally, we measured participants' cognitive load during the study. While Lindlbauer et al.~\cite{lindlbauer2019context} employed the three MR activities as requiring different cognitive loads in application scenarios, they did not empirically demonstrate these differences. To address this gap, we administered the NASA-TLX questionnaire~\cite{hart1988development}, which provided a standardised assessment of participants' perceived cognitive workload across the different activities. 

\paragraph{Apparatus}
We utilised a Windows 11 laptop with an NVIDIA GeForce RTX 4080, an Intel Core i9-13980HX processor, and 32GB RAM, which connected to the Meta Quest 3 headset. This configuration supported the pass-through functionality required for our MR environment. The software application was developed and implemented using Unity version 2022.3.22f1.

\subsection{Results}

\begin{figure}[b]
    \centering
    \includegraphics[width=\linewidth]{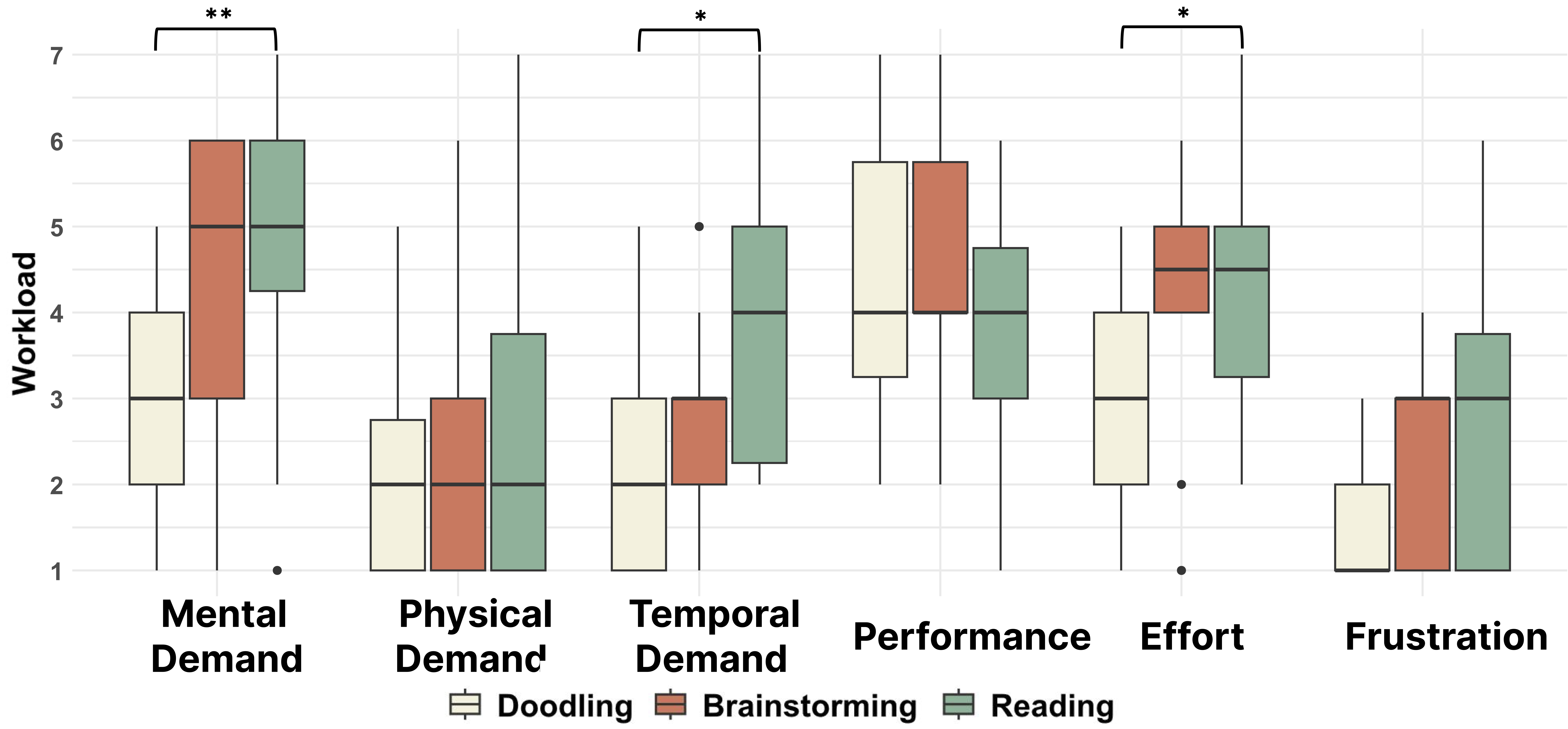}
    \caption{The raw NASA-TLX results ($p < .05$(*), $p < .01$(**).}
    \label{fig:NASA}
    \vspace{-6mm}
\end{figure}

The NASA-TLX analysis revealed significant differences in cognitive workload across the three conditions (Doodling, Brainstorming, and Reading). Shapiro-Wilk tests confirmed normal distribution only for Performance across activities (Doodling: $p = .283$, Brainstorm: $p = .316$, Reading: $p = .243$) with no significant differences found via one-way repeated measures ANOVA ($p = .190$). Friedman tests for non-normally distributed dimensions revealed significant differences in Mental Demand ($\chi^2(2) = 13.82, p = .001$), Temporal Demand ($\chi^2(2) = 9.97, p = .007$), Effort ($\chi^2(2) = 8.27, p = .016$), and Frustration ($\chi^2(2) = 10.68, p = .005$). Post-hoc analyses with Bonferroni correction demonstrated that Reading induced significantly higher Mental Demand than Doodling ($p = .004$), as well as higher Temporal Demand ($p = .013$) and required more Effort ($p = .025$). No significant differences were found for Physical Demand ($p = .491$). For overall workload, which met normality assumptions, Repeated Measures ANOVA revealed significant differences between activities ($p<.001$). Post-hoc analysis indicated that Brainstorming required significantly higher overall workload than Doodling ($p=.041$), while Reading demanded even more significantly ($p=.004$). No significant difference was found between Reading and Brainstorming ($p=.169$). Our analysis confirms that the three MR activities successfully created varied cognitive demands, with reading imposing the highest workload, followed by Brainstorming and Doodling, validating our experimental design for studying notification behaviour across different cognitive states. These findings support our methodology by confirming we collected data across meaningfully different contexts.
\subsection{Implications}
Our user study establishes two essential datasets: a self-labelled dataset ($N=18\times90$) and an interaction-based dataset ($N=18\times108$) capturing notification behaviour during three activities with varying cognitive demands. Our NASA-TLX analysis confirms these activities successfully created distinct cognitive workload conditions, with Reading imposing a significantly higher overall workload than Doodling, validating our experimental design and ensuring data collection across meaningfully different cognitive states. This methodological foundation directly supports our subsequent analyses in \autoref{sec:human_behaviour} by collecting users' self-reported behaviour patterns through the ESM. The confirmed differences in cognitive load across activities will be particularly valuable when analysing how Activity influences notification responsiveness. Furthermore, these datasets provide the necessary training and testing data for developing our personalised notification urgency classifier (PersoNo) in~\autoref{sec: Classifier}, where we will evaluate different classification approaches using both self-labelled and interaction-based data to determine which yields the most effective urgency predictions.

\section{Study 2: Human Behaviour Patterns}
\label{sec:human_behaviour}

\begin{table}[t]
\centering
\caption{Codebook for Notification Response Patterns in Mixed Reality. The frequency of each code assignment is indicated in parentheses. Individual notifications may receive multiple codes.}
\label{tab:codebook}
\Large
\resizebox{1\linewidth}{!}{ 
\begin{tabular}{p{60pt}p{120pt}p{200pt}}
\toprule
\textbf{Theme} & \textbf{Sub-Theme} & \textbf{Definition} \\
\midrule
\multirow{4}{*}{Sender (11)} & \cellcolor{gray!20} Authority-Based Prioritisation (8) &  \cellcolor{gray!20}Preferential response to notifications from supervisors\\
 & Social Relationship Prioritisation (3) & Preferential response to notifications from friends \\
 &  \cellcolor{gray!20} Group Message Ignorance (8) &  \cellcolor{gray!20} Tendency to ignore group messages \\
\hline
\multirow{4}{*}{Content (14)} & Action Request Response (12) & Tendency to respond to notifications requiring action or questions \\
 &  \cellcolor{gray!20} Content Length Sensitivity (5) &  \cellcolor{gray!20} Response patterns influenced by notification length \\
 & Information Density Evaluation (3) & Response based on perceived information value  \\
 &  \cellcolor{gray!20} Implicit Content Cues (3) &  \cellcolor{gray!20} Response influenced by implicit cues of notification  \\
\hline
\multirow{4}{*}{Activity (14)} & Cognitive Load Management (4) & Response patterns based on cognitive demands of current activity \\
 &  \cellcolor{gray!20} Activity Engagement Level (2) &  \cellcolor{gray!20} Response patterns influenced by engagement with current activity \\
 & Activity-Specific Response Strategies (14) & Different response strategies for different MR activities \\
 &  \cellcolor{gray!20} Task Disinterest Displacement (3) &  \cellcolor{gray!20}Higher response rate due to disinterest in primary task \\
\bottomrule
\end{tabular}}
\vspace{-5mm}
\end{table}

While existing research established theoretical frameworks for mobile notifications~\cite{li2023alert, pielot2014situ, mehrotra2016myPhoneAndMe}, our work adopts a top-down thematic approach~\cite{braun2006thematic} to systematically examine how these frameworks manifest differently in MR. The unique perceptual and contextual factors of MR, where digital information overlays physical space, likely transform how users perceive, prioritise, and respond to notifications. It raises our \textbf{RQ1: How do users behave and respond to notifications in MR?} 
The next contribution is understanding human behaviour patterns regarding MR notifications. Our novel insights establish the critical contextual variables that must be prioritised when developing MR notification classifiers, directly informing our later prompt construction methodology. 

\subsection{Thematic Analysis}
Two researchers independently reviewed all users' ESM data and developed the initial codebook, referencing previous work by~\cite{li2023alert, mehrotra_prefminer_2016}. We identified the Sender, Content, and Activity as the most frequently mentioned factors, which formed the basis of three thematic categories. After independently drafting initial codes, we collaboratively refined definitions and examples through discussion. Subsequently, we conducted a pilot phase on two self-reported behaviour patterns ($\sim10\%$ of the data) to test and refine the codebook.

Following codebook construction, we randomly selected four users' behavioural patterns ($\sim22\%$ within the range suggested by O'Connor et al.~\cite{o2020intercoder}) to validate our codebook, ensuring these were distinct from those used in the pilot phase. In addition to the original researchers who developed the codebook, we invited another researcher to code the user behaviour using the established codebook. We calculated Krippendorff's alpha, a standard inter-rater reliability measure for non-mutually exclusive coding schemes~\cite{hayes2007answering}. Our Krippendorff's alpha was 0.846 above the standard of reliable labelling results~\cite{hayes2007answering}. After validating our codebook's reliability, two researchers independently coded all users' ESM data and subsequently resolved any coding disagreements. \autoref{tab:codebook} presents the defined codebook and the frequency of code assignment.

\subsection{Results}

Our analysis identified three primary themes influencing notification response in MR: Sender, Content, and Activity contexts (see \autoref{tab:codebook}). Beyond these main themes, participants reported additional notification response patterns, such as opportune timing (N=2). Two participants indicated that opportune timing is crucial in determining whether they would reply to messages within 30 seconds. Interestingly, \textit{P15} mentioned responding to notifications somewhat randomly, even when the content was important, to avoid giving others the impression of constant availability. We categorised these as ``Others'' and excluded them from our analysis since only a few participants mentioned them.

Participants’ likelihood of attending to a notification depended on who the sender was (N=11), but not always in the way seen with smartphones. Surprisingly, personal friendship played a minimal role in the immediate MR notification responses. Only three instances were recorded. This differs markedly from traditional mobile notification research~\cite{pielot2014situ, chang2023not}, where personal relationships significantly influence notification attendance. Instead, participants gave preferential attention to notifications from their supervisors (N=8), while group notifications were often ignored (N=8).

Content characteristics emerged as equally important, with action request responses (N=12) dominating this theme. Participants also demonstrated sensitivity to content length (N=5) and employed sophisticated information evaluation through implicit cues (N=3) and information density assessment (N=3). While Li et al.~\cite{li2023alert} found content factors influenced mobile notification preferences more than contextual factors, our results indicate that in MR, content considerations maintain equivalent importance alongside sender and activity contexts, rather than dominating them.

Activity context featured prominently, with activity-specific response strategies (N=14) representing our most frequent code. It indicates that users often developed different notification response rules for various activities in MR. Cognitive load management (N=4) and task disinterest displacement (N=3) revealed how participants balanced attention resources. Two participants reported that they would use the Mute All function in reality when engaging in any activities and respond to the notifications during breaks.

\subsection{Implications}
\label{sub: implication}
Our thematic analysis establishes a comprehensive framework that advances MR notification classifier development by identifying key themes (Sender, Content, and Activity) and their critical sub-themes affecting user responsiveness. MR designers can use our codebook to develop more sophisticated classification systems incorporating nuanced factors.

It is noticeable that users' MR notification behaviour patterns differ from their self-reported receptivity to mobile notifications (see \autoref{sub:participant}). A substantially higher number of participants considered both Sender and Activity factors when deciding whether to respond to MR notifications. This finding suggests that Activity represents a crucial variable when designing an MR notification classifier, even though Li et al.~\cite{li2023alert} demonstrated that utilising notification content alone for mobile notification classifiers could achieve reasonable results. The discrepancy highlights the unique contextual considerations necessary for effective notification management in MR compared to traditional mobile settings.

Furthermore, our findings reveal a significant departure from traditional mobile notification patterns. While previous research suggests users typically prioritise responses to friends' messages due to social pressure~\cite{chang2023not, pielot2014situ}, only a few participants reported their priorities in social relationships. Instead, users overrode this factor with other considerations. Additionally, the substantially higher number of participants who adopted activity-oriented notification response strategies may be attributed to the more disruptive nature of MR notifications, resulting in reduced notification receptivity during different activities. Thus, for the following notification classifier development, in addition to our hypothesised variables (Sender and Content), we also considered the Activity.

\section{Study 3: Notification Classifier}
\label{sec: Classifier}
\begin{figure}[t]
    \centering
    \includegraphics[width=\linewidth]{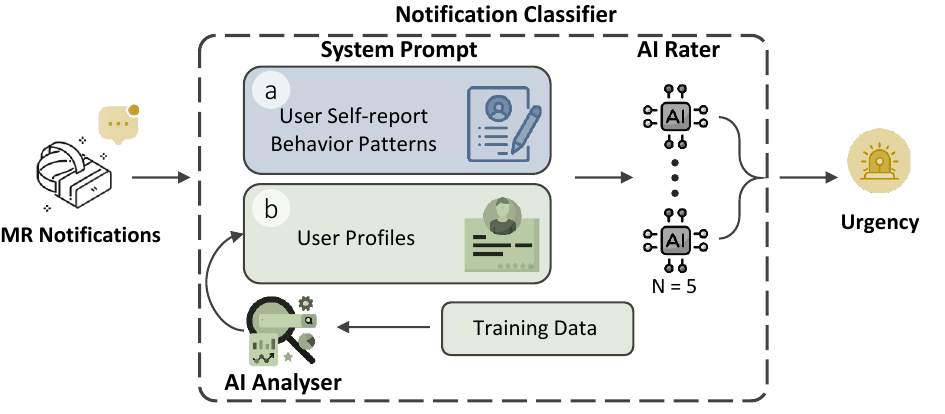}
    \vspace{-4mm}
    \caption{Framework of the Classifiers with different In-Context Learning Prompts. \circled{a} shows \textbf{M$_1$} zero-shot learning using user self-reported behaviour patterns. \circled{b} refers to the \textbf{M$_2$} multi-agent methods that an AI analyser extracts user profiles from the training data.}
    \vspace{-4mm}
    \label{fig:classifier_framework}
\end{figure}

We developed an intelligent notification management system with the training dataset (Study 1) and the key contextual factors identified in Study 2. Next, we describe our notification classifiers and evaluate hypotheses related to three essential classifier components: Data (\textbf{[H2]}), Context (\textbf{[H3]}), and Algorithm (\textbf{[H1]}, \textbf{[H4]}).

\textbf{[H1]}: Personalised notification classifier will outperform the general notification classifier. \textbf{[H2]}: Self-reported urgency ratings will yield classification performance comparable to interaction-based data. \textbf{[H3]}:  Activity, mentioned as frequently as Content by our participants, will outperform classification performance beyond models using only Sender and Content variables. \textbf{[H4]}: LLM analysers will capture user notification response patterns accurately and comprehensively based on our codebook framework.

\subsection{Data Preparation}
\label{sub:Classifier/Data Preparation}
We utilised part of the notification data collected during MR activities as the test data. Following previous work~\cite{li2023alert}, we designated the last six notifications from each activity as testing data ($N=6\times3=18$). This methodology emulates real-world scenarios where historical data informs predictive models of human behaviour patterns. For analysis, we prepared three distinct training datasets: \textbf{SR}: Self-Report, containing data labelled directly by participants during the self-report phase; \textbf{D$_1$} (Dataset 1), an interaction-based notification dataset collected during MR activities with key variables of Sender and Content; and \textbf{D$_2$} (Dataset 2), identical to D$_1$ but incorporating Activity as an additional contextual variable, which ~\autoref{sub: implication} identified as crucial.

\subsection{Model Design}
Unlike previous research on notification classifier~\cite{chen2022predicting, mehrotra2015designingContentDriven} that implemented traditional Machine Learning (ML) algorithms, we leverage LLMs to build our notification classifier. In our study, we utilised Qwen~\cite{qwq32b} (\href{https://huggingface.co/Qwen/QwQ-32B}{QwQ-32B}), an open-source LLM, to ensure reproducibility. Pretrained on large-scale corpora, LLMs have demonstrated essential language understanding capabilities~\cite{brown2020language}. This foundation enables the model to analyse the semantic meaning of notification content and capture general urgency indicators, such as calls to action and time-sensitive content~\cite{li2023alert}. Furthermore, previous work~\cite{brown2020language} has shown that In-Context Learning (ICL) enables pre-trained language models to more effectively address downstream tasks, in our case, notification urgency classification. This technique requires a small amount of data, making LLMs ideally suited for personalized notification classification, which inherently has limited training data since all the data are from a single user. Additionally, ICL provides an interpretable interface for LLM interaction~\cite{brown2020language}, which enhances the explainability of the algorithm. Compared with traditional ML approaches, using LLMs with ICL is more aligned with standards of Human-Centered Artificial Intelligence (HCAI)~\cite{shneiderman2020human}, prioritising user understanding and control.

Three models were included in our study. We employed two ICL methods to optimise the LLMs for notification urgency classification: (\textbf{M$_1$}) Zero-shot learning~\cite{brown2020language} utilising only user-reported behaviour patterns collected via ESM, and (\textbf{M$_2$}) Multi Agent (MA)~\cite{wangself} implementing an analyser LLM to create user profiles from training data, which rater LLMs then use to classify test data. Apart from the personalised classifiers using M$_1$ and M$_2$, we evaluated general notification classifiers using base models (\textbf{Base}) to predict the test dataset directly. These models, trained on general corpora, analyse urgency levels from the general users' perspective.

To further improve the classification reliability and accuracy, we integrated homogeneous LLM self-ensemble techniques~\cite{wangself} (see \autoref{fig:classifier_framework}) and Chain-of-Thoughts (CoT) prompting~\cite{wei2022chain}. Each method employed five raters with a temperature setting of 1, resulting in five independent votes per notification with certain randomness. It was specifically employed to reduce prediction variance across multiple raters. The final urgency label was determined by majority vote. Rather than direct classification, CoT prompted LLMs to articulate their reasoning process step-by-step before making predictions, significantly improving transparency and accuracy.

\subsection{Prompt Design}
\label{sub: prompt}
We designed two ICL prompts as the base prompts based on D$_1$/SR and D$_2$. We augmented these base prompts with specific information to construct complete prompts for each methodological test, such as ``The following is the user behaviour pattern: \{user\_pattern\}.'' While being used for the Base, we left the \{user\_pattern\} blank. The prompts hold the variables in different datasets: the first prompt (\textbf{P1}) instructed the model to analyse only the \textit{Sender} and \textit{Content} variables, suitable for SR and D$_1$, while the second prompt (\textbf{P2}) expanded the analysis to include \textit{Sender}, \textit{Content}, and \textit{Activity} variables, corresponding to D$_2$. These prompts guided models in formulating reasoning based on the respective variables before generating predictions. Further, for the analyser LLMs in the MA framework, we integrated findings from the thematic analysis (see \autoref{sec:human_behaviour}), instructing the analyser to consider all identified sub-themes when developing comprehensive user profiles. We applied a similar method to design the rater prompts.

\subsection{Model Configurations}
\label{sub:Classifier/Model Configurations}

We prompted models using factorial combinations of methods and datasets $\textsc{Method}\times\textsc{Dataset}$. For example, with $\textsc{M$_2$}\times\textsc{D$_1$}$, we employed the multi-agent (M$_2$) method and prompted analyser models with interaction-based data containing Sender and Content variables (D$_1$) using P1-based prompts. More specifically, M$_1$ operates without a training notification dataset, relying exclusively on user-reported behavioural patterns. Consequently, we utilised both prompts (P1 and P2, detailed in \autoref{sub: prompt}) to evaluate this method, with P1 analysing Sender and Content variables while P2 incorporated Activity as an additional contextual factor. Additionally, we evaluated a general notification classifier (Base) that, similar to M$_1$, employed both P1 and P2 prompts but without any information about user profiles. These Base models relied solely on their pre-training on general corpora to analyse notification urgency from a non-personalised perspective.

\subsection{Results}

\begin{table}[t!]
    \vspace{-3mm}
    \centering
    \small
    \setlength{\tabcolsep}{4pt}
    \caption{Comparison of models}
    \resizebox{1\linewidth}{!}{ 
\begin{tabular}{p{50pt}p{20pt}p{20pt}p{20pt}p{20pt}p{20pt}p{20pt}p{20pt}}
        \toprule
        & \multicolumn{2}{c}{\textbf{Base}} & \multicolumn{2}{c}{\textbf{M$_1$}} & \multicolumn{3}{c}{\textbf{M$_2$}} \\
        \cmidrule(r){2-3} \cmidrule(lr){4-5} \cmidrule(l){6-8}
        & \multicolumn{1}{c}{D$_1$} & \multicolumn{1}{c}{D$_2$} & \multicolumn{1}{c}{D$_1$} & \multicolumn{1}{c}{D$_2$} & \multicolumn{1}{c}{SR} & \multicolumn{1}{c}{D$_1$} & \multicolumn{1}{c}{D$_2$} \\
        \midrule
        Accuracy & 0.670 & 0.670 & 0.735 & 0.679 & 0.747 & 0.759 & \textbf{0.815} \\
        FNR & 0.703 & 0.627 & 0.528 & 0.614 & 0.432 & 0.550 & \textbf{0.381}\\
        Specificity & 0.887 & 0.846 & 0.838 & 0.818 & 0.821 & \textbf{0.923} & 0.875 \\
        AUROC & 0.595 & 0.609 & 0.684 & 0.617 & 0.721 & 0.708 & \textbf{0.786} \\
        \bottomrule
    \end{tabular}}
    \vspace{-5mm}
    \label{tab: Classifier_results}
\end{table}

\begin{figure}[b!]
    \centering
    \includegraphics[width=\linewidth]{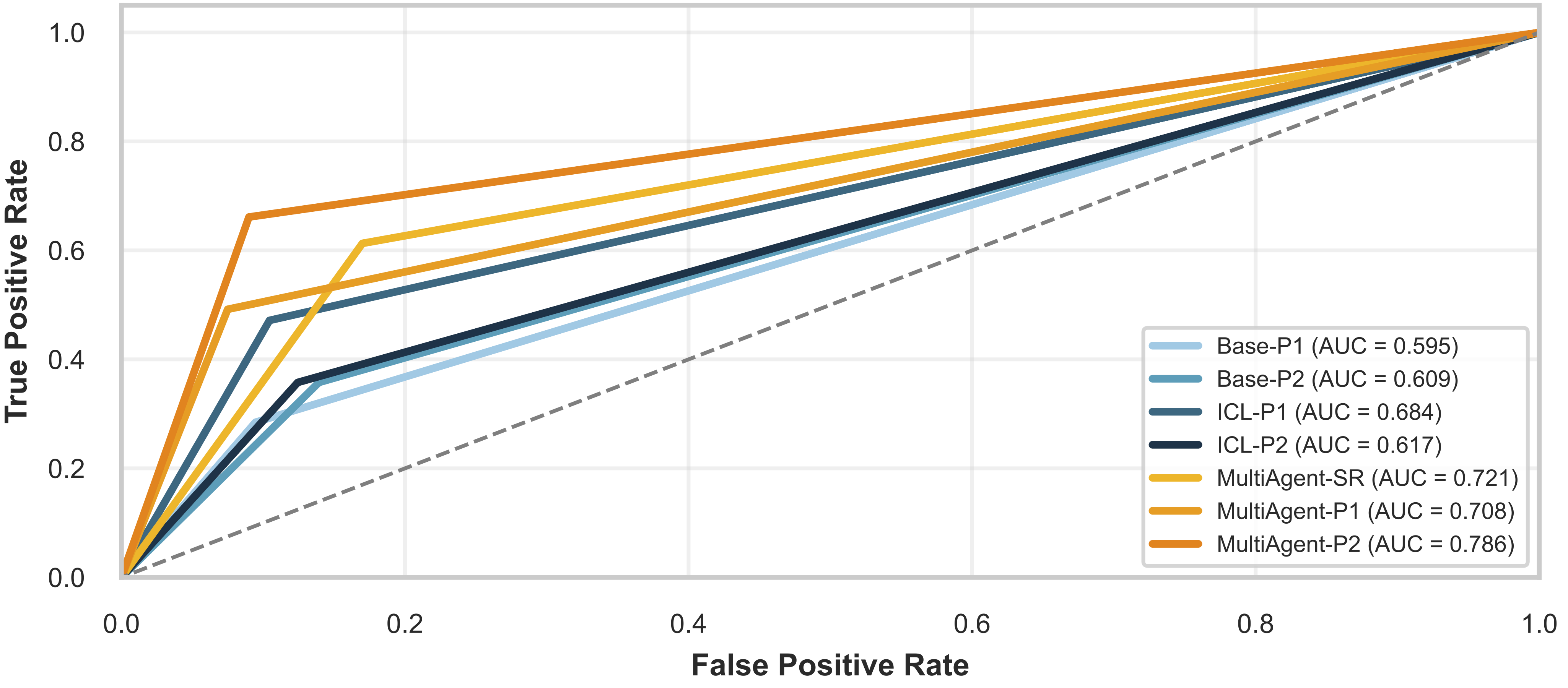}
     \vspace{-4mm}
    \caption{AUROC Results. TPR represents the True Positive Rate, and FPR refers to the False Positive Rate.}
    \label{fig:AUROC}

\end{figure}

We evaluated model performance using Accuracy, False Negative Rate (FNR), Specificity, and Area Under the Receiver Operating Characteristic (AUROC). While accuracy measures overall correctness, it can be misleading with imbalanced notification urgency classes. Specificity (true negative rate) measures the system's ability to filter out non-urgent notifications, directly addressing our goal of reducing unnecessary interruptions in MR. Furthermore, we incorporated the FNR to quantify missed urgent notifications. Based on previous work's findings~\cite{iqbal2010notificationsDefinition, mehrotra2016myPhoneAndMe} that users tolerated interruptions to avoid missing important updates, we assumed that the cost of missing important notifications was higher than the cost of being disturbed and prioritised minimising FNR. AUROC (see \autoref{fig:AUROC}) provides a threshold-independent assessment of discriminative capability, remaining robust against the common imbalance where non-urgent notifications typically outnumber urgent ones. 

See \autoref{tab: Classifier_results} for the performance of all model configurations. To evaluate our hypotheses regarding Data (\textbf{[H2]}) and Context (\textbf{[H3]}), we primarily focused on comparing \textit{PersoNo} (M$_2$), our proposed notification system, which demonstrated superior performance across all metrics compared to alternative models. Statistical significance between conditions was assessed using pairwise t-tests.

For \textbf{[H1]}, we compared personalised notification classifiers against the general classifier (Base). Results showed that personalised approaches (M$_1$ and M$_2$) consistently outperformed the base model across all metrics. The base model achieved only 0.670 accuracy with P1 and P2, while personalised models reached up to 0.815 accuracy with M$_2$ on D$_2$. Notably, the FNR of the base model was substantially higher (0.586 and 0.523) than personalised approaches, indicating that users would be more likely to miss important information using the general classifier.

For \textbf{[H2]}, we examined whether self-reported urgency ratings yield comparable performance to interaction-based data. The M$_2$ model using SR achieved 0.747 accuracy and 0.721 AUROC, while the interaction-based data (D$_2$) resulted in 0.759 accuracy and 0.708 AUROC. Pairwise t-tests showed no significant differences between these approaches in accuracy and FNR ($t(17)=.334$, $p=.742$ and $t(17)=1.99$, $p=.066$), supporting our hypothesis that self-reports serve as viable alternatives for classifiers.

\begin{figure}[t!]
    \centering
    \includegraphics[width=\linewidth]{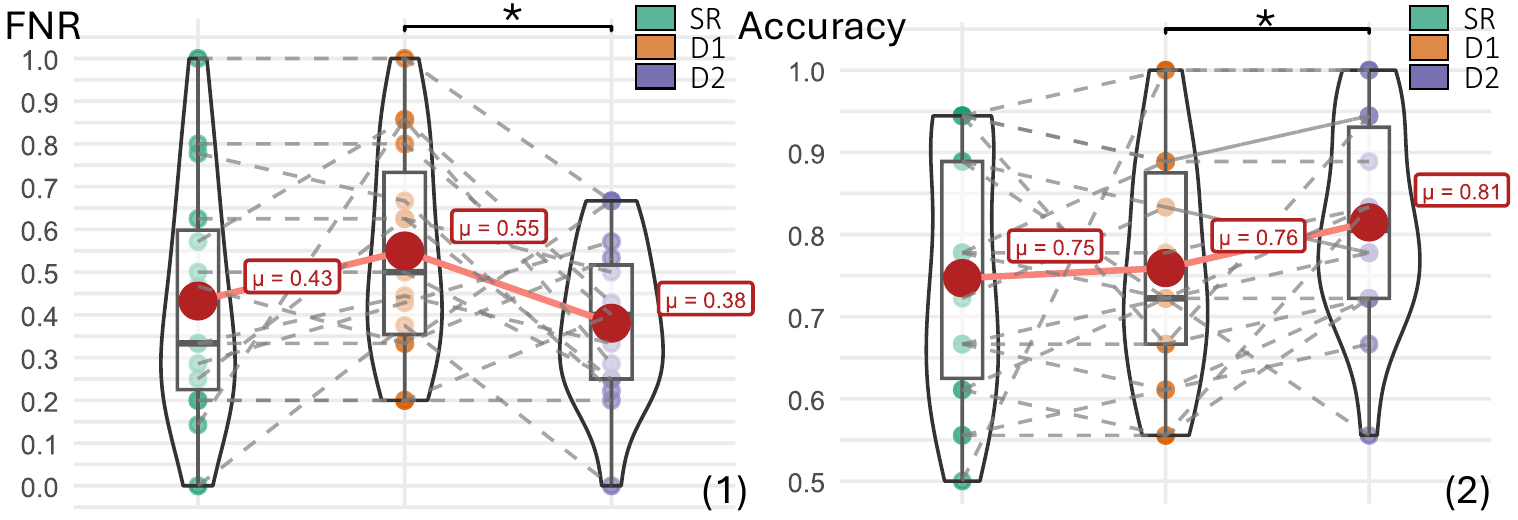}
    \caption{Performance comparison of M$_2$ (\textit{PersoNo}). (1) False Negative Rate and (2) Accuracy across conditions. \textcolor{grey}{Dashed lines} indicate changes in individual participant results ($p < .05$(*)).}
    \vspace{-4mm}
    \label{fig:MA_results}
\end{figure}

For \textbf{[H3]}, we investigated whether incorporating Activity context would improve classification performance. The model trained on D$_2$ (which included Activity context) significantly outperformed D$_1$ (Sender and Content only) in both FNR (0.381 compared to 0.550, $t(17)=2.30$, $p=.037$) and accuracy (0.815 compared to 0.759, $t(17)=-2.15$, $p=.046$). Moreover, D$_2$ demonstrated enhanced AUROC (0.786 compared to 0.708), confirming that Activity represents a crucial contextual factor in MR notification management. ~\autoref{fig:MA_results} (1) demonstrates a reduction in FNR for the majority of participants, as indicated by the general downward trend of the gray dashed lines, while (2) exhibits the opposite pattern, revealing an improvement in Accuracy.

For \textbf{[H4]}, we assessed whether LLM analysers could effectively capture user notification response patterns. The superior performance of M$_2$, consisting of LLM analysers generating user profiles and raters providing predictions, validated this hypothesis. M$_2$ consistently outperformed M$_1$ prompted by P2, with the D$_2$ configuration achieving significantly higher accuracy (0.815, $t(17)=3.335$, $p=.004$) and lower FNR (0.381, $t(17)=-3.008$, $p=.009$).

\subsection{Implications}

Analysis of our experimental results provides substantial evidence supporting all four hypotheses. However, we found some unexpected decreases in the model performance with the incorporation of the Activity context. \autoref{tab: Classifier_results} shows the accuracy of M$_1$ decreased from 0.735 to 0.679 when LLM raters were prompted to consider Activity context, which appears to contradict our \textbf{[H3]}.

Several factors can explain this apparent contradiction. First, the ESM data revealed that not all participants reported activity-based behavioural patterns regarding MR notifications. In this context, incorporating the Activity component would likely complicate the rating process and potentially bias final evaluations by introducing redundant information. Second, participants' self-reported activity-oriented behavioural patterns were often expressed in broad terms that lacked the specificity needed for fine-grained classification. This generality made it difficult for LLM raters to detect subtle differences between individual users' notification preferences across various activities. When LLMs were instructed to consider Activity without sufficient user-specific data, they likely defaulted to general population assumptions rather than personalised predictions. For instance, LLMs might automatically classify notifications during Reading as non-urgent based on its typical high cognitive load, regardless of individual user preferences. However, some participants reported difficulty focusing on the Reading task due to a lack of interest, which increased their willingness to respond to IMs. Furthermore, the zero-shot learning approach employed in M$_1$ appears particularly sensitive to this issue, as it lacks training examples that would help calibrate activity-based predictions to individual users.

Nevertheless, the more sophisticated M$_2$ approach, which incorporated explicit training examples, successfully leveraged Activity context to improve classification performance, ultimately supporting \textbf{[H3]} within more structured learning frameworks.

\section{Discussion}
\subsection{Human-Centered PersoNo}
The \textit{PersoNo} system, comprising an LLM analyser and raters, demonstrated superior performance (accuracy of 0.815 and FNR of 0.381 with M$_2$-D$_2$), highlighting the potential of LLM applications in personalised systems. Furthermore, the interpretability of LLM-based systems represents a significant advantage. As stated by Shneiderman~\cite{shneiderman2020human}, HCAI should not only focus on the algorithmic performance but also empower users by offering control and understanding. Unlike black-box ML models, our CoT approach made the classification process transparent, with LLM raters explicitly articulating their reasoning before making predictions. This transparency improved classification accuracy and could also enhance user trust and system adoption. Additionally, our multi-agent architecture enhances the system's human-centered qualities by separating profile generation from notification classification. This separation enables potential user intervention in profile creation, allowing users to review and adjust automatically extracted preferences. By aligning fundamental aspects of HCAI design, such capability affords users direct control over how the system interprets their notification behaviour, fostering transparency and autonomy.

\subsection{Impacts and Applications of PersoNo}
\paragraph{Advancing Notifications Management from Mobile to MR}

Our work extends mobile intelligent notification systems research~\cite{li2023alert, mehrotra_prefminer_2016} to the unique challenges of MR environments. While prior mobile notification classifiers such as PrefMiner~\cite{mehrotra_prefminer_2016} and content-driven systems~\cite{mehrotra2015designingContentDriven} established foundational approaches for managing interruptions, they operated under fundamentally different interaction paradigms. Mobile users can physically distance themselves from devices, but MR notifications directly overlay the user's visual field.
PersoNo addresses this gap by introducing the personalised notification classifier specifically designed for MR's spatial computing context. Our finding that activity context equals content importance (see  \autoref{sec:human_behaviour}) gives different perspectives from Li et al.'s~\cite{li2023alert} well-known mobile-centric framework, where content dominated other factors. This shift reflects the fundamental difference in how notifications compete for cognitive resources in MR, a finding that extends Lindlbauer et al.'s~\cite{lindlbauer2019context} work on context-aware MR interfaces to the notification domain.

\paragraph{Redefining Personalisation Via Limited Data Learning}
Traditional personalised notification systems required extensive data collection periods. For example, Mehrotra et al.~\cite{mehrotra_prefminer_2016} needed 15 days, while Pielot et al.~\cite{pielot2017beyond} averaged four weeks. This requirement has been a significant barrier to adoption, as noted by Maister~\cite{maister1984psychology}. PersoNo fundamentally reconceptualises the notification classification through LLM-based learning, achieving 81.5\% accuracy with just 90 training instances (\textbf{[H4]}). It demonstrates the superior performance with a limited dataset. 
This advancement builds upon recent work in few-shot learning~\cite{brown2020language} but applies it specifically to the HCI challenge of notification management. Our multi-agent architecture (M$_2$) represents a novel application of LLM capabilities to extract meaningful user profiles from minimal data—a contribution that extends beyond notification systems to any personalised intelligent interface requiring rapid adaptation to individual users.

\paragraph{Bridging VR/MR Notification Design and MR Intelligence}

While VR and MR notification research has advanced placement strategies~\cite{hsieh2020bridging, rzayev2019notification} and multimodal design~\cite{ghosh2018notifivr}, these studies primarily addressed notification presentation rather than intelligent filtering. PersoNo bridges this critical gap by introducing personalised urgency classification, which enables dynamic adaptation of established notification design principles. Our system transforms static design guidelines into context-aware behaviours: urgent notifications leverage the bottom-center placement proven most noticeable~\cite{rzayev2018reading, plabst2022push}, while non-urgent messages can utilise less intrusive in-situ positioning~\cite{rzayev2019notification} that preserves spatial context without demanding immediate attention. In addition to the adaptive notification placement, the future VR/MR developers may consider pushing the non-urgent notifications until the opportune time, such as the break during VR/MR activities~\cite{chen2022predicting}.

This urgency-based adaptation directly supports the vision of calm technology in MR~\cite{a2w}, where information should inform without overwhelming. By filtering notifications before they reach the presentation layer, PersoNo ensures that only contextually appropriate interruptions utilise prime visual real estate. Furthermore, the system's classification output can further integrate with advanced MR adaptation frameworks like SituationAdapt~\cite{Li2024SituationAdapt}, which employs vision-and-language models for environmental analysis. While PersoNo determines notification urgency based on user patterns, SituationAdapt could identify optimal spatial placement by analysing the user's current visual scene, creating a comprehensive pipeline from urgency assessment to context-aware positioning.

Our current implementation employs binary urgency classification, aligning with established notification research methodologies~\cite{mehrotra2015designingContentDriven}. However, the modular architecture of our approach facilitates future extensions to multi-level urgency schemes through prompt-based redefinition of urgency categories. Such granular classification would unlock the full potential of existing VR/MR notification placement~\cite{rzayev2019notification}, enabling nuanced placement strategies where notification position, opacity, and persistence vary along an urgency continuum rather than a simple binary threshold.
\subsection{Subjective Labelling and Objective User Behaviour}
Our study compared two data collection approaches for training notification classifiers: self-labelled urgency ratings (subjective) and actual interaction behaviour (objective). Although both datasets were employed in previous notification systems~\cite{chen2022predicting, pielot2017beyond}, prior subjective data were collected primarily through ESM. Consequently, the literature lacks analysis of user-labelled datasets and their comparison with actual interaction behaviour datasets. The comparable performance between models trained on self-reported data (M$_2$-SR) and interaction-based data (M$_2$-D$_1$) suggests that self-reporting can effectively substitute for longer-term interaction tracking when building personalised notification classifiers (\textbf{[H2]}). This finding contradicts prior work suggesting weak correlations between self-reporting and actual behaviours~\cite{dang2020self}. The unexpected consistency between subjective and objective data may be due to our study's focus on the specific interaction behaviour of notification responses. Users are highly familiar with notification patterns in their everyday messaging interactions and are likely to possess a strong awareness of both their notification preferences and behavioural tendencies. 

Thus, rather than extended data collection periods spanning weeks, which has been standard practice in previous notification research~\cite{mehrotra_prefminer_2016, oh2015intelligent}, personalised notification systems could be deployed using simple user-provided labels. In addition to PersoNo's ability to classify notification urgency accurately with limited personal data, this approach could significantly improve user acceptance of intelligent notification systems by minimising the waiting period before deployment. Beyond the notification, a potential implication is that all intelligent systems could leverage subjective feedback as training data for interactions with users who are highly familiar with them. We acknowledge that human behaviour patterns vary over time~\cite{mehrotra_prefminer_2016, mehrotra2015designingContentDriven}. However, we propose using self-labelled data to initiate and facilitate user adoption of the personalised system. Such implementations can continuously adapt based on users' interaction data to accommodate changing behaviour patterns.

\subsection{Limitations}
Our study establishes foundational insights for MR notification management while revealing opportunities for future research. First, our self-labelling methodology focused on content and sender variables, following established mobile notification research paradigms. While this approach successfully demonstrated the viability of self-reported data for classifier training, future work could enhance this methodology by incorporating activity-specific rating scenarios to capture the full contextual richness we identified as crucial for MR environments. 
Second, our controlled laboratory environment and pre-established social relationships (using participant-provided names as placeholders) ensured systematic variable manipulation but may not fully capture real-world MR notification dynamics. In-the-wild deployments with a longitudinal study would provide valuable insights into PersoNo's performance under authentic conditions and evolving social relationships. 
Finally, while our study identified three key contextual dimensions (content, sender, activity) with 18 participants, notification receptivity in MR likely depends on additional factors. Future work could explore broader contextual variables, including temporal factors (time of day)~\cite{sarker2014assessing}, spatial contexts (location~\cite{mehrotra2015designingContentDriven}, proximity to others), and MR-specific factors (immersion level, virtual-physical task integration), while validating findings with larger sample sizes.

\section{Conclusion}
Our research addresses notification management in MR by introducing \textit{PersoNo}, a personalised LLM-based notification urgency classifier. 
Through user studies, we collected the first MR notification dataset and discovered that activity context is equally important as content and sender in MR, which is a key difference from mobile notification management. 
\textit{PersoNo} achieved 81.5\% accuracy, 0.786 AUROC, and 0.381 FNR by effectively analysing user profiles from limited data, outperforming baseline approaches. 
Notably, with \textit{PersoNo}, self-reported urgency ratings proved effective for classifier training while considering the contexts, contradicting assumptions about self-report validity. 
To conclude, adhering to HCAI design principles, \textit{PersoNo} employs AI to minimise distractions while ensuring notification awareness, simultaneously providing users with understanding and control.

\acknowledgments{
This research was supported by the Hong Kong Polytechnic University's Start-up Fund for New Recruits (No. P0046056), Departmental General Research Fund (DGRF) from HK PolyU ISE (No. P0056354), and PolyU RIAM -- Research Institute for Advanced Manufacturing 
(No. P0056767). Jingyao Zheng and Xian Wang were supported by a grant from the PolyU Research Committee under student account codes RMCU and RMHD, respectively.}
This work has been partly supported by the Research Center Trustworthy Data Science and Security (\href{https://rc-trust.ai}{https://rc-trust.ai}), one of the Research Alliance centers within the UA Ruhr (\href{https://uaruhr.de}{https://uaruhr.de}).


\bibliographystyle{abbrv-doi-narrow}

\bibliography{main}
\end{document}